# Multi-spectral piston sensor for co-phasing giant segmented mirrors and multi-aperture interferometric arrays

**François Hénault**
UMR 6525 CNRS H. Fizeau – UNS, OCA Avenue Nicolas Copernic, 06130 Grasse – France

E-mail: francois.henault@obs-azur.fr

**Abstract.** This paper presents the optical design of a multi-spectral piston sensor suitable to co-phasing giant segmented mirrors equipping the Future Extremely Large Telescopes (ELTs). The general theory of the sensor is described in detail and numerical simulations have been carried out, demonstrating that direct piston and tip-tilt measurements are feasible within accuracies respectively close to 20 nm and 10 nano-radians. Those values are compatible with the co-phasing requirements, although the method seems to be perturbed by uncorrected atmospheric seeing.

## 1 Introduction

From the Hooker telescope of 100" diameter build during the 1920s on Mount Wilson to the achievement of the Very Large Telescope (VLT) array in Chile, the 20th century has unquestionably demonstrated the superiority of large reflective telescopes in the field of astronomical observations. It is commonly believed, however, that the classical operations of manufacturing, polishing and supporting large glass mirrors will soon be confronted to their technological limits, and that in view of 10-m class (or higher) ground-based telescopes, the primary mirrors will need to be composed of several smaller individual reflective facets (or segments), a major choice having been validated on the two 10-m Keck telescopes. For space observatories alternatively, the mirror diameters are rather limited by the space available under the cone of the launching rocket, leading to a current maximum below 4 meters. Hence the James Webb Space Telescope (JWST, to be operated in 2014) will be equipped with a 6.5-m segmented mirror, being deployable and optically adjustable in space.

In the case of a giant segmented mirror, all the reflective facets must be individually adjusted in piston (along a direction parallel to the telescope optical axis) and tip-tilt (rotations around two axes perpendicular to the telescope optical axis) so that the assembled segments ideally mimic the theoretical, continuous surface of the mirror. This operation is sometimes called the "co-phasing" of the telescope and must be carried out within a given accuracy, which could be $\lambda/4$ Peak-to-Valley (PTV) or $\lambda/13.4$ Root Mean Square (RMS) according to either the Rayleigh or Maréchal criteria, where $\lambda$ is the wavelength of the electro-magnetic field. In this paper is chosen a target accuracy of $\lambda/10$ RMS, which is frequently quoted in papers relevant to the co-phasing of telescopes and sparse apertures interferometers (see for example [1-2]), and constitutes a reasonable magnitude order at least when coronagraphic applications are not envisaged. For a telescope of diameter 5 m, this requirement corresponds to approximate piston and tip-tilt tolerances of respectively 25 nm and 10 nano-radians in the visible range (λ = 0.5 µm). Those requirements are extremely demanding and necessitated to develop specific alignment techniques, such as those summarized below:
- The first pioneering works were undeniably undertaken at the Keck telescope: Chanan *et al* [3-4] upgraded some already well-known Wavefront Sensors (WFS) concepts, such as the Shack-Hartmann or curvature WFS, in order to give to them the capacity to discriminate piston errors: here it has to be noticed that, since their basic principle consists in measuring the local slopes or curvatures of the wavefronts (WFE) before reconstructing them digitally, those WFS are not naturally well-suited for piston sensing. It was then shown that enhanced hardware and algorithms searching for local slope or curvature discontinuities can be used to determine the piston errors. However these methods are hardly applicable to diluted apertures interferometers, a condition that we seek to satisfy in this study.
- Another option is to employ phase retrieval or phase diversity digital procedures, since the latter have already been applied successfully to the determination of phase errors on both monolithic telescopes and multi-aperture optical systems [5-7]. However the technique usually requires significant post-processing times, which prevents them from being operated on-ground in an adaptive optics (AO) regime: thus an "ideal" WFS should indeed combine the ability to perform direct WFE measurements in quasi real-time, which adds another stringent requirement to the system.
- Some alternative WFS concepts based on a Mach-Zehnder interferometer (or on an equivalent principle) installed at the focal plane of the optical system to be co-phased have also been proposed by different authors [8-10], but none of them seem to have been validated on-sky.
In this paper we finally choose to re-examine the concept of a multi-wavelength, phase-shifting "Telescope-Interferometer" (TI), another focal plane WFS that has been described and studied recently in its monochromatic version [11-13]. Its principle, making use simple numerical algorithms could allow to quickly and directly evaluate the WFEs (including piston errors) created on either giant segmented mirrors or multi-aperture interferometers. We first recall briefly the monochromatic theory of the TI in section 2.1, before extending it to the case of multiple wavelengths in section 2.2. A tentative optical design based on the combination of a phase-shifting stage and a multi-spectral stage is then described in section 3. An end-to-end numerical model intended to evaluate the WFS performance is briefly described in section 4 and its preliminary results in terms of piston and tip-tilt measurement accuracy are presented, before a short summary is provided in section 5.

## 2    Theory

In this section is first recalled the principle of WFE measurements performed using a monochromatic phase-shifting TI, before generalizing it to the polychromatic case. Basically, the proposed technique consists in adding a second, reference optical beam into the main pupil in order to generate modulated and phase-shifted point spread functions. The searched phase errors can then be extracted from demodulating the Fourier transforms of the PSFs in the Fourier plane. More details about the TIs and their theory can be found in Refs. [11-13].

### *2.1    The monochromatic Phase-Shifting Telescope Interferometer (PSTI)*

Let us consider a telescope of 5-m diameter whose primary mirror is constituted of N individual segments disposed following an hexagonal arrangement as depicted in Figure 1. It is assumed that the central segment does not exist, and is replaced by a smaller, circular reference mirror of diameter d = 2r centred on point O. Named "reference pupil" in the remainder of the text, it is supposed this mirror can be displaced longitudinally along the Z optical axis, thereby introducing a phase-shift $\phi_m$ of the central reference pupil with respect to the other mirror segments ($\phi_m$ will further take M different values, with $1 \leq m \leq M$, see below). The expression of the complex amplitude $A_m(P)$ in the exit pupil plane OXY therefore writes (see Figure 1):

$$A_m(P) = B_r(P)\exp[i\phi_m] + \sum_{n=1}^{N} B_D(P - P_n)\exp[ik\Delta_n(P - P_n)], \qquad (1)$$

where $B_r(P)$ is the amplitude transmitted by the reference pupil (here equal to the "pillbox" or "top-hat" function of radius r), $B_D(P)$ is the two-dimensional amplitude transmission function of the hexagonal segments, $P_n$ is the centre of the $n^{th}$ segment, $\Delta_n(P)$ is the WFE to be measured and $k = 2\pi/\lambda$. In this section will only be considered the piston and tip-tilt errors $\xi_n$ and $\mathbf{t_n}$ of the segments, hence:

$$\Delta_n(P - P_n) = \xi_n + \mathbf{t_n}\,\mathbf{P_n P}, \tag{2}$$

where $\mathbf{t_n}$ stands for the unitary vector perpendicular to the $n^{th}$ facet. In the frame of scalar diffraction theory, the complex amplitude distribution $\hat{A}_m(M')$ in the telescope image plane O'X'Y' is equal to the Fourier transform of $A_m(P)$:

$$\hat{A}_m(M') = \hat{B}_r(M')\exp[i\phi_m] + \hat{B}_D(M')\sum_{n=1}^{N}\exp[ik(\xi_n + \mathbf{t_n}\,\mathbf{P_n P})]\exp[-ik\mathbf{OP_n}\,\mathbf{O'M'}/F], \tag{3}$$

where $\hat{B}_r(M')$ and $\hat{B}_D(M')$ respectively are the Fourier transforms of $B_r(P)$ and $B_D(P)$, and F is the focal length of the segmented telescope. The Point-Spread Function (PSF) of the system is by definition equal to the square modulus of $\hat{A}_m(M')$, i.e. after multiplying with its complex conjugate:

$$\begin{aligned}PSF_m(M') = &\ PSF_r(M') + PSF_T(M') \\ &+ \exp[-i\phi_m]\big[\hat{B}_r\,\hat{B}_D\big](M')\sum_{n=1}^{N}\exp[ik(\xi_n + \mathbf{t_n}\,\mathbf{P_n P})]\exp[-ik\mathbf{OP_n}\,\mathbf{O'M'}/F] \\ &+ \exp[i\phi_m]\big[\hat{B}_r\,\hat{B}_D\big]^*(M')\sum_{n=1}^{N}\exp[-ik(\xi_n + \mathbf{t_n}\,\mathbf{P_n P})]\exp[ik\mathbf{OP_n}\,\mathbf{O'M'}/F],\end{aligned} \tag{4a}$$

where $PSF_r(M')$ and $PSF_T(M')$ respectively stand for the PSFs of the reference pupil and of the whole segmented telescope:

$$PSF_r(M') = \big|\hat{B}_r(M')\big|^2 \tag{4b}$$

$$\begin{aligned}PSF_T(M') = \big|\hat{B}_D(M')\big|^2 &\left\{\sum_{n=1}^{N}\exp[ik(\xi_n + \mathbf{t_n}\,\mathbf{P_n P})]\exp[-ik\mathbf{OP_n}\,\mathbf{O'M'}/F]\right\} \\ &\times\left\{\sum_{n=1}^{N}\exp[-ik(\xi_n + \mathbf{t_n}\,\mathbf{P_n P})]\exp[ik\mathbf{OP_n}\,\mathbf{O'M'}/F]\right\}.\end{aligned} \tag{4c}$$

One of the two basic principles of the PSTI now consists in physically measuring the point spread functions $PSF_m(M')$ on a CCD camera (or another type of detector array), and in computing digitally their associated Optical Transfer Functions (OTF) by means of an inverse Fourier transform. From Eq. (4), one finds an expression of $OTF_m(M')$ that is composed of four different terms:

$$\begin{aligned}OTF_m(P) = &\ OTF_r(P) + OTF_T(P) \\ &+ \exp[-i\phi_m]\sum_{n=1}^{N}\exp[ik(\xi_n + \mathbf{t_n}\,\mathbf{P_n P})]\big[B_r * B_D\big](P - P_n) \\ &+ \exp[i\phi_m]\sum_{n=1}^{N}\exp[-ik(\xi_n + \mathbf{t_n}\,\mathbf{P_n P})]\big[B_r * B_D\big](P + P_n),\end{aligned} \tag{5}$$

with symbol * denoting a convolution product. The second essential point of the PSTI procedure is then to isolate the third term of Eq. (5) by means of an appropriate set of phase-shifts $\phi_m$ allowing to linearly combine all the computed OTFs:

$$\begin{aligned}C(P) = \frac{1}{M}\sum_{m=1}^{M}OTF_m(P)\exp[i\phi_m] = &\ \frac{1}{M}\left\{\sum_{m=1}^{M}\exp[i\phi_m]\right\}[OTF_r(P) + OTF_T(P)] \\ &+ \sum_{n=1}^{N}\exp[ik(\xi_n + \mathbf{t_n}\,\mathbf{P_n P})]\big[B_r * B_D\big](P - P_n) \\ &+ \frac{1}{M}\left\{\sum_{m=1}^{M}\exp[2i\phi_m]\right\}\sum_{n=1}^{N}\exp[-ik(\xi_n + \mathbf{t_n}\,\mathbf{P_n P})]\big[B_r * B_D\big](P + P_n),\end{aligned} \tag{6}$$

and it is found that the second term can be easily separated from the others if the selected phase-shifts are satisfying the both conditions:

$$\sum_{m=1}^{M}\exp[i\phi_m] = \sum_{m=1}^{M}\exp[2i\phi_m] = 0. \tag{7}$$

Figure 2 provides a graphic illustration of possible solutions for M = 3 and M = 4 in the complex plane – it has to be noticed that only the quadruplet $\{\phi_1;\phi_2;\phi_3;\phi_4\} = \{0;\pi/2;\pi;3\pi/2\}$ was originally considered in the previous publications [11-13]. When the conditions (7) are respected the final expression C(P) of the linearly combined OTFs becomes:

$$C(P) = \sum_{n=1}^{N} \exp[ik(\xi_n + \mathbf{t_n}\ \mathbf{P_n P})] [B_r * B_D](P - P_n) \approx \sum_{n=1}^{N} \exp[ik(\xi_n + \mathbf{t_n}\ \mathbf{P_n P})] B_D(P - P_n), \quad (8)$$

and, if the spatial dimensions of the reference pupil are significantly smaller than those of the hexagonal facet, the function $B_r(P)$ can be replaced by the Dirac distribution $\delta(P)$, and the phase of C(P) can be considered as constant over the whole segment area, being simply proportional to the searched piston $\xi_n$. The justification and implications of this approximation have been extensively discussed in Ref. [14]. The final step of the PSTI procedure consists in a phase extraction that is carried out by isolating the phase errors of the $n^{th}$ segment, multiplying both sides of Eq. (8) with the function $B_D(P-P_n)$:

$$C(P)\ B_D(P - P_n) = \exp[ik(\xi_n + \mathbf{t_n}\ \mathbf{P_n P})]\ B_D(P - P_n), \quad (9)$$

and the estimated phase simply writes:

$$\varphi_n = (\xi_n + \mathbf{t_n}\ \mathbf{P_n P})/\lambda = \arctan\{\operatorname{Im}[C(P)\ B_D(P - P_n)]/\operatorname{Real}[C(P)\ B_D(P - P_n)]\}/2\pi. \quad (10)$$

All the previous steps are schematically summarized by the flow-chart of Figure 3. The overall performance of the method was studied in more detail in Refs.[11-12], leading among others to the following conclusions.

- The intrinsic measurement accuracy of the PSTI is better than $\lambda/5$ PTV and $\lambda/100$ RMS, which makes it competitive with respect to the best existing WFS. However in monochromatic light the measurement range stays limited by the so-called "$2\pi$ ambiguity" – i.e. the phase can only be retrieved modulo $2\pi$ from Eq. (10).
- The observed light source does not need to be purely monochromatic: maximal spectral bandwidths $\delta\lambda/\lambda$ of 20% can be tolerated without significant loss of performance.
- Although the method is particularly efficient in space, it may also be operated on ground in an adaptive optics regime: depending on the seeing conditions and on the size of the reference pupil, it can accept guide stars up to the magnitude 11 in the V band.

Next section is now devoted to the extrapolation of this monochromatic WFE measurement method to several different wavelengths and spectral bands in order to extend its capture range beyond the $2\pi$ ambiguity.

### 2.2 Multi-wavelength operation

Extending the measurement range of the Optical Path Differences (OPD) between different separated telescopes is a common problem in multi-aperture interferometry that can be solved by combining phase measurements simultaneously performed in different spectral bands. Such methods can indeed be extrapolated to the measurement of piston errors and to the co-phasing of large segmented mirrors. For example, we describe in the Appendix how to adapt the "dispersed speckles" method recently proposed by Borkowski *et al* [2][15-16] to the case of a polychromatic PSTI. It turns out, however, that the required number of spectral channels is probably too important, and would be detrimental to the phase measurement accuracy (that was shown to be inversely proportional to the spectral width of an individual measurement channel [12]). Hence the following study will be restricted to methods that make use of a minimal number of wavelengths, inspired from the techniques of multi-colour phase shift interferometry [17-18] and non-contact length and distance metrology [19-20]. Given an optical distance to be determined $\xi_n$, the basic principle consists in combining its fractional phases $\varphi_{ln}$ measured for L different wavelengths $\lambda_l$, with $1 \leq l \leq L$. Limiting the total wavelength number to L = 3, the problem reduces to solving the following equations system:

$$\xi_n = (n_{1n} + \varphi_{1n})\lambda_1 = (n_{2n} + \varphi_{2n})\lambda_2 = (n_{3n} + \varphi_{3n})\lambda_3 \quad (11)$$

The latter is actually a 3N system of 4N unknown variables, which are $\xi_n$ and the positive or negative integer numbers $n_{1n}$, $n_{2n}$ and $n_{3n}$. Due to the integer nature of $n_{ln}$, this under-constrained system can nevertheless be solved over a limited domain of piston values $[-\lambda_S/2, \lambda_S/2]$, where $\lambda_S$ is defined as the "synthetic wavelength". Herein we select an heuristic expression of $\lambda_S$, which still belongs to Tilford's original family of solutions [19]:

$$\lambda_S = \left(1/\lambda_1 - 2/\lambda_2 + 1/\lambda_3\right)^{-1}, \qquad (12)$$

assuming that $\lambda_1 < \lambda_2 < \lambda_3$. This enables to define a simple multi-spectral OPD unwrapping procedure whose pseudo-code is provided in Figure 4 and major steps are described below (here the indices n have been omitted for the sake of clarity):

1) Starting from the fractional phases $\varphi_1$, $\varphi_2$ and $\varphi_3$ acquired for the three different wavelengths, a first guess of the piston error is estimated as $\delta_0 = \lambda_S (\varphi_1 - 2\varphi_2 + \varphi_3)$, and $\delta_0$ is further brought back into the [-$\lambda_S$/2, $\lambda_S$/2] range if necessary.
2) First guess $\delta_0$ and the fractional phases together allow to estimate the integer numbers $n_1$, $n_2$ and $n_3$.
3) Three improved estimations of the piston error $\delta_1$, $\delta_2$ and $\delta_3$ can now be derived from those integer numbers and the fractional phases.
4) The final piston error $\xi$ is defined as the arithmetical mean of $\delta_1$, $\delta_2$ and $\delta_3$.
5) A "piston measurement accuracy index" noted $\sigma_\delta$ and equal to the variance of $\delta_1$, $\delta_2$ and $\delta_3$ is also computed, serving as a quality estimator for the obtained result.

The fifth and last step of the numerical procedure is perhaps the most important, since it may be comprehended as a sanity check of the whole measurement sequence: errors in step n°4 can originate either from the intrinsic uncertainty $\delta\varphi$ affecting the measured fractional phases, or from false estimations of $n_1$, $n_2$ or $n_3$. In the latter case, the final measurement uncertainty $\sigma_\delta$ should be at least equal to $\lambda_1/3$. If on the contrary the three integer values are unbiased, $\sigma_\delta$ can be estimated as:

$$\sigma_\delta = \delta\varphi/3 \left[\lambda_1^2 + \lambda_2^2 + \lambda_3^2\right]^{1/2} \approx \lambda_2\, \delta\varphi/\sqrt{3}, \qquad (13)$$

and it is therefore possible to define a simple criterion (e.g. $\sigma_\delta < \lambda_1/10$) allowing to accept or reject the current estimation of the piston error.

To conclude, it must be emphasized that the major advantages of the here above presented multi-spectral OPD unwrapping procedure are the fact that it only involves very simple mathematical relationships suitable to real-time computing, on the one hand, and the possibility of a self sanity check allowing to reject numerical results corrupted by real measurement errors, on the other hand. The following section now aims at defining a tentative optical layout for the envisaged multi-wavelength, phase-shifting piston sensor.

## 3    Description of the piston sensor

The theoretical considerations that have been exposed in the previous section can serve as a starting point for designing the preliminary optical architecture of a WFS based on the principle of the multi-spectral PSTI. The major requirements may be summarized as follows:

1) This is basically a focal plane wavefront sensor, whose volume and dimensions near the telescope focus (or one of its images) should be kept as small as possible in order to ensure good mechanical and thermal stabilities.
2) The WFS shall include a phase-shifting device located at an image plane of the segmented mirror to be characterized (assumed to be the entrance pupil of the telescope).
3) The WFS shall provide the capacity of acquiring simultaneously or within a very short lap of time the PSFs of the telescope in several narrow spectral channels.

Two very preliminary designs answering to the two first requirements have already been described in Ref. [13]: in the first one the beams are divided by an arrangement of M-1 beamsplitters and the PSFs are acquired simultaneously on M different, synchronized detector arrays. In the second configuration all the PSFs are measured sequentially on one single camera within a total acquisition time not exceeding 10 msec. In addition to the advantage of requiring only one detector array, the latter solution was found the most favourable in terms of radiometric performance, especially when photon noise dominates detector noise (i.e. for bright sky objects). In the herein presented design, the phase-shifts $\phi_m$ are introduced sequentially while the measurements in different spectral bands are simultaneous. It is assumed that the primary mirror of the telescope includes a "reference segment" of superior image quality (i.e. diffraction-limited), corresponding to the reference pupil area where the phase-shifts are introduced (two examples of practical implementation of the reference pupil are provided in Ref. [11]). The optical layout of the piston sensor is schematically presented in Figure 5: it is essentially composed of three main sub-systems, namely the phase-shifting stage, the multi-spectral stage and the CCD camera that are described below. Numerical simulations of the performance of the complete optical system are further presented in section 4.

*3.1     Phase-shifting stage*
This sub-system is essentially composed of one collimating lens imaging the telescope entrance pupil (i.e. the giant segmented mirror) on a reference flat mirror that is pierced at the location of the telescope reference pupil. A mandrel carrying a small, flat and high-precision optical surface is piezo-electrically shifted along the optical axis, thus generating the required successive phase-shifts $\phi_m$, for all indices m comprised between 1 and M. However, many other designs could certainly be envisaged owing to the recent progress of Micro-opto-electromechanical Systems (MOEMS) technology.

*3.2     Multi-spectral stage*
The core function of the multi-spectral stage is to perform simultaneous measurements of $PSF_m(M')$ – as defined in Eq. (4a) – in different channels of moderated spectral width. The proposed design makes use of an image/pupil inversion that was originally suggested by Courtès and Viton for the observation of gaseous nebulae and distant galaxies [21]. It consists in setting a transmissive diffraction grating near an intermediate image plane X'Y', spreading the beams in different directions depending on their wavelengths, and to re-arrange them in the spectrally dispersed pupil plane XY (see the middle part of Figure 5, which is not to scale). This can be achieved by means of a few optical designer tricks:
- The employed diffraction grating is engraved on a convergent optical element (here represented as a lens) in order to re-imaging the dispersed entrance pupil on its slicer with high demagnification ratio, therefore allowing to define spectral channels $\delta\lambda_l$ with sharp edges (it must be noticed that although the diffraction grating presented in Figure 5 is a transmissive one, any reflective grating may be used as conveniently).
- The wavelength separation is realized by a "pupil slicer" located at the dispersed output pupil plane, also serving to redirect the beams towards different areas of the detector array. This pupil slicer can be made of a micro-lens array associated to a converging lens as represented in Figure 5, a design that was already manufactured and test for the instrument MUSE of the VLT [22]. The pupil slicer could also be fully reflective: the realization of such mini arrays of optical elements is nowadays well mastered (see for example Ref. [23]).
- Finally, the focal lengths of the individual elements of the pupil slicer are not identical, but inversely proportional to the wavelengths $\lambda_l$ in order to compensate for the linear magnifications of the measured functions $PSF_m(M')$ with respect to $\lambda_l$. The real necessity of this wavelength rescaling has been confirmed by the numerical simulations presented in the next section. The rescaling might also be performed digitally (e.g. using interpolation algorithms) rather than optically, however it has been chosen here to simplify the data processing software as much as possible, in order to stay compatible with real time applications.

*3.3     CCD camera*
This final stage of the piston sensor consists in a high performance detector array, either of the CCD or CMOS type, typically characterized by high quantum efficiency and low read-out noise over the considered spectral domain, and equipped with a 12-bit analogue-to-digital converter.

## 4     Numerical simulations

In this section are presented the numerical results of an end-to-end optical model (described in section 4.1) intended to assess the expectable performance of the described WFS in terms of phase and pistons measurement errors. For a given set of telescope and instrumental parameters (section 4.2), the simulated cases cover static piston and tip-tilts errors (section 4.3), the influence of the selected spectral bandwidths $\delta\lambda_l$ (section 4.4), and the effects of atmospheric disturbance for a ground-based sensor (section 4.5).

*4.1 The numerical model*

The numerical model of the multi-wavelength piston sensor is actually based on the monochromatic model that was described in Ref. [11]. It is indeed split into two major modules, respectively simulating the $PSF_m(M')$ functions recorded by the camera, and subsequently computing the OTFs via inverse Fourier transforms and applying the whole data processing defined by Eqs. (8-12). The multi-spectral OPD unwrapping procedure of Figure 4 is entirely included into the second module. The first module is basically a ray-tracing software having the ability of modelling different apertures of various shape and sizes, taking into account alignment errors in both piston and tip-tilt and atmospheric seeing. This instrument simulator then computes the monochromatic PSFs and sums them incoherently over the considered spectral channels. The program can also add various types of detection noises and digitization errors to the PSFs, even if the latter possibility was not utilized in the frame of this study. All these operations are repeated for the M phase-shifts $\phi_m$ and the L different spectral channels centred on the wavelengths $\lambda_l$.

*4.2 Definition of the system and input parameters*

We consider a telescope of diameter 5 m and focal length 50 m, hence being open at F/10. It is composed of N = 18 hexagonal facets arranged as shown in Figure 1 (this configuration was originally intended to mimic the JWST. Those parameters would of course need to be updated in the case of the future ground-based ELTs, although it is likely that the major conclusions would not change significantly). For all simulations, it is assumed that the phase-shifted, reference pupil is centred on the optical axis of the telescope, which should not hamper the conclusions. Random piston errors ranging from –10 µm to +10 µm are added to each telescope segment, as well as tip-tilt errors comprised between –1 and +1 micro-radian around the X and Y-axes. Those pistons and tip-tilts values are supposed to represent realistic alignment errors of the segments resulting from the employed pre-alignment technique (for comparison purpose, they are twice more optimistic than the initial errors measured on the Keck 2 telescope [3]). All the numerical values are provided in Table 1 and the global WFE map of the segmented mirror is depicted by the grey-scale map and the perspective view of Figure 6.

**Table 1: Initial, measured and difference values of the piston and tip-tilt errors for a spectral bandwidth $\delta\lambda/\lambda = 10\%$ and a reference pupil radius r=500 mm.**

| Segment number | Initial errors | | | Measured errors | | | Differences | | |
|---|---|---|---|---|---|---|---|---|---|
| | Piston (µm) | Tilt X (µrad) | Tilt Y (µrad) | Piston (µm) | Tilt X (µrad) | Tilt Y (µrad) | Piston (µm) | Tilt X (µrad) | Tilt Y (µrad) |
| 1 | -3.837 | 0.633 | -0.487 | -3.859 | 0.646 | -0.498 | -0.023 | 0.013 | -0.012 |
| 2 | 6.584 | 0.024 | 0.190 | 6.584 | 0.025 | 0.201 | 0.000 | 0.001 | 0.011 |
| 3 | 0.714 | 0.156 | -0.286 | 0.709 | 0.163 | -0.288 | -0.006 | 0.007 | -0.002 |
| 4 | -5.000 | -0.517 | -0.304 | -4.990 | -0.515 | -0.310 | 0.011 | 0.002 | -0.006 |
| 5 | -6.063 | -0.708 | -0.631 | -6.062 | -0.711 | -0.641 | 0.001 | -0.003 | -0.009 |
| 6 | -4.141 | -0.856 | 0.570 | -4.136 | -0.876 | 0.588 | 0.005 | -0.021 | 0.017 |
| 7 | 0.950 | -0.514 | 0.862 | 0.950 | -0.521 | 0.878 | 0.000 | -0.006 | 0.016 |
| 8 | 3.571 | 0.439 | 0.096 | 3.555 | 0.432 | 0.098 | -0.015 | -0.007 | 0.002 |
| 9 | 3.023 | -0.367 | 0.784 | 3.029 | -0.374 | 0.800 | 0.006 | -0.007 | 0.016 |
| 10 | -8.079 | -0.251 | -0.480 | -8.066 | -0.261 | -0.486 | 0.012 | -0.009 | -0.006 |
| 11 | 2.669 | 0.989 | -0.921 | 2.702 | 1.005 | -0.935 | 0.033 | 0.016 | -0.013 |
| 12 | -3.693 | 0.393 | 0.424 | -3.708 | 0.401 | 0.433 | -0.015 | 0.009 | 0.009 |
| 13 | 3.831 | -0.577 | 0.154 | 3.829 | -0.593 | 0.164 | -0.002 | -0.016 | 0.010 |
| 14 | -3.409 | 0.727 | 0.020 | -3.401 | 0.729 | 0.017 | 0.008 | 0.002 | -0.003 |
| 15 | 1.728 | -0.985 | 0.753 | 1.690 | -0.988 | 0.766 | -0.038 | -0.002 | 0.013 |
| 16 | 6.855 | -0.742 | -0.624 | 6.881 | -0.764 | -0.644 | 0.026 | -0.023 | -0.019 |
| 17 | -5.472 | 0.008 | 0.364 | -5.481 | 0.008 | 0.373 | -0.010 | 0.001 | 0.009 |
| 18 | 4.688 | -0.317 | 0.027 | 4.678 | -0.314 | 0.014 | -0.010 | 0.003 | -0.013 |
| Average | -0.282 | -0.137 | 0.028 | -0.283 | -0.139 | 0.029 | -0.001 | -0.002 | 0.001 |
| PTV | 14.934 | 1.974 | 1.783 | 14.947 | 1.992 | 1.813 | 0.071 | 0.039 | 0.037 |
| RMS | 4.685 | 0.587 | 0.537 | 4.686 | 0.595 | 0.547 | 0.017 | 0.011 | 0.012 |

In order to minimize the total time requested for the acquisition of the M phase-shifts $\phi_m$, we choose M = 3 and consequently $\{\phi_1;\phi_2;\phi_3\} = \{0;2\pi/3;4\pi/3\}$. The selection of the number of wavelengths L, of the central wavelengths $\lambda_l$ and of the spectral widths $\delta\lambda_l$ of each spectral channel is not a straightforward task, and was here carried out in a iterative manner, trying to establish a compromise between the following adverse tendencies:

- The piston capture range increases as more wavelengths are added and their values are closed one to the other. However, this criterion tends to reduce the spectral widths $\delta\lambda_l$.
- For radiometric reasons, the measurement accuracy increases with $\delta\lambda_l$, but this is detrimental to the capture range and augments the risk of failure of the OPD retrieval algorithm (the point is further discussed in section 4.4).

We finally adopted L = 3 and the triplet $\{\lambda_1;\lambda_2;\lambda_3\} = \{0.456;0.502;0.552\}$ µm. The other numerical parameters are provided in Table 2. For such values of the three central wavelengths, the synthetic wavelength $\lambda_S$ is equal to 48.75 µm according to the relation (12), which yields a capture range of the piston errors being theoretically equal to [-24.375,+24.375] µm, largely compliant with their actual figures. Moreover, setting M = L = 3 allows to limit the total number of axial displacements of the reference mirror in the phase-shifting stage to seven, being respectively equal to 0 (common to all spectral channels), and the triplets $\{\lambda_1/3;\lambda_2/3;\lambda_3/3\}$ and $\{2\lambda_1/3;2\lambda_2/3;2\lambda_3/3\}$ in terms of OPD. The total computational load is essentially dominated by the seven required Fast Fourier Transform (FFT) operations: assuming WFE map sampling of 512 x 512 as in the present study, it is found that quasi-real time operation requires approximately 20 GFlops, a performance that is easily fulfilled by modern laptop computers.

**Table 2: Numerical parameters of the three selected spectral channels.**

| Central wavelength (µm) | Maximal spectral bandwidth $\delta\lambda/\lambda$ (µm) | | Effective focal length (m) |
|---|---|---|---|
| | | (%) | |
| $\lambda_1$    0.456 | 0.044 | 9.6 | 55.04 |
| $\lambda_2$    0.502 | 0.048 | 9.6 | 50. |
| $\lambda_3$    0.552 | 0.052 | 9.4 | 45.47 |

*4.3   Piston and tip-tilt sensing capacity*

Let us firstly consider the case of static piston and tip-tilt errors resulting from a preliminary alignment of the segmented primary mirror. We set the maximal spectral bandwidth $\delta\lambda/\lambda$ of the three measurement channels to 10 % and the reference pupil radius to its maximal possible value, i.e. r = 500 mm (top left panel of Figure 7). The numerical results of the simulation are provided in Table 2 and illustrated in Figure 7, showing the uncorrected PSF (middle row) and the retrieved WFE at the surface of the exit pupil (bottom left panel), as well as its difference with respect to the original errors (bottom right panel). It must be noticed that the numerical values provided in Table 2 result from linear regressions of the WFE map at the surface of each individual segment, limited to a 700 mm useful diameter circle in order to maximize the piston measurement quality estimator $\sigma_\delta$ defined in section 2.2. It is found that the piston and tip-tilt estimation errors are finally equal to 17 nm and 11/12 nano-radians (with respect to the X/Y axes) in RMS sense, a satisfactory result that is nearly compliant with the original goals defined in section 1.

*4.4   Influence of spectral bandwidth*

As was already mentioned in section 4.2, The WFE measurement accuracy is expected to improve with the spectral width of the three measurement channels, on the one hand, but this benefit is counterbalanced by a smaller capture range and the possibility of false estimations of the integer numbers $n_1$, $n_2$ and $n_3$ (see § 2.2), on the other hand. The latter can be evaluated quantitatively by means of an "OPD retrieving confidence ratio" defined as follows:

$$\rho = N_\delta / N_T \tag{14}$$

where $N_T$ is the total number of measurement points at the surface of an individual mirror segment, and $N_\delta$ is the number of points for which the criterion $\sigma_\delta < \lambda_1/10$ is fulfilled ($\sigma_\delta$ being defined by Eq. (13) in § 2.2). For each segment of the previous simulated case, the evolution of $\rho$ as a function of the relative spectral bandwidth $\delta\lambda/\lambda$ is tabulated in Table 3 and graphically illustrated in Figure 8, where the confidence ratio is encoded into a grey scale: white areas are those where the sanity check $\sigma_\delta < \lambda_1/10$ was found successful, while other tones progressively darken as the discrepancies are increased (black tone representing error values superior or equal to 1 µm). It can be seen that the "safe areas" where piston errors can be unambiguously determined progressively shrink as $\delta\lambda/\lambda$ tends to its maximal value of 10%, possibly leading to low confidence ratios such as 18% and 9% for segments n°9 and 10 respectively. It is quite remarkable however that correct piston estimations can still be achieved, probably making advantage of the high number $N_T$ of available measurement points and of the self sanity check inherent to the method. Another conclusion derived from the presented numerical results is that the worst confidence ratios are not only associated to large piston errors, but also to the initial tip-tilt alignment errors of the segments, and this effect stays noticeable even for small piston errors (see for example segments n° 7, 11 and 15).

**Table 3: Influence of the spectral bandwidth $\delta\lambda/\lambda$ on the OPD retrieving confidence ratio for each mirror segment.**

| Segment number | Spectral bandwidth $\delta\lambda/\lambda$ | | | |
| --- | --- | --- | --- | --- |
| | 3% | 5% | 7.5% | 10% |
| 1 | 100 | 100 | 100 | 100 |
| 2 | 100 | 100 | 100 | 100 |
| 3 | 100 | 100 | 59 | 30 |
| 4 | 100 | 100 | 46 | 23 |
| 5 | 100 | 100 | 100 | 78 |
| 6 | 100 | 100 | 100 | 100 |
| 7 | 100 | 100 | 100 | 64 |
| 8 | 100 | 100 | 100 | 91 |
| 9 | 100 | 49 | 25 | 18 |
| 10 | 100 | 100 | 29 | 9 |
| 11 | 100 | 99 | 39 | 24 |
| 12 | 100 | 100 | 100 | 100 |
| 13 | 100 | 100 | 100 | 100 |
| 14 | 100 | 100 | 100 | 100 |
| 15 | 100 | 95 | 35 | 19 |
| 16 | 100 | 100 | 100 | 60 |
| 17 | 100 | 100 | 100 | 100 |
| 18 | 100 | 100 | 100 | 100 |

### 4.5 Influence of atmospheric perturbations

Keeping the same piston and tip-tilt static errors as above, we finally add moderate seeing perturbations being characterized by a Fried's radius $r_0 = 500$ mm [24], and try to recover the initial wavefront and geometrical alignment errors. It can then be readily observed that the method breaks down when significant reference pupil dimensions or spectral bandwidths are simulated by the numerical model. As an example, the Figure 9 illustrates the results obtained for a reference pupil radius $r = 200$ mm and $\delta\lambda/\lambda = 7.5$ %, showing the perturbed WFE (top right panel), the resulting PSF (middle left panel, logarithmic scale), the reconstructed WFE (middle right panel) and a raw difference map between both WFEs (bottom left panel). It has been noticed that most of the residual error originates from one particular segment (the n°10, where the confidence ratios $\rho$ were found lower, see Table 3). Removing this segment from the main pupil allows to greatly improve the global measurement accuracy (see the bottom right panel of Figure 9), however this stays insufficient for entering within the requirements. It can be concluded that atmospheric perturbations severely limit the capacities of the method, which cannot be employed for disentangling the piston errors from the seeing. Hence the method should probably be restricted to spatial applications, or to ground telescopes already equipped with an adaptive optics system, or with a mono-mode WFE filtering subsystem such as single-mode fibers or integrated optics [25].

## 5 Conclusion

In this paper was presented the optical design of a multi-spectral piston sensor suitable for co-phasing giant segmented mirrors equipping the future Extremely Large Telescopes (ELTs). The general theory of the sensor has been presented in detail and numerical simulations demonstrated that direct piston and tip-tilt measurements are feasible within accuracies respectively close to 20 nm and 10 nano-radians in RMS sense. Those values are compatible with the usual co-phasing requirements, although the method is severely perturbed by uncorrected atmospheric seeing. It must be emphasized that, although it was presented herein in the sole framework of the ELT, this method is fully applicable to sparse aperture interferometers.

**Appendix. Connection with the "dispersed speckles" method**

The dispersed speckles method was originally proposed by Borkowski *et al* for the co-phasing of ground or space multi-aperture interferometers [2][15]. It basically consists in reorganizing multi-spectrally measured PSFs on a three-dimensional data-cube, whose third dimension is scaled in terms of the wavenumber σ (the inverse of the wavelength λ). A three-dimensional inverse Fourier transform of the data-cube then allows to recover information about the piston errors along the third vertical axis. However the floor map of this data-cube – being described by relation (4b) – often presents a complex structure, making it difficult to retrieve the original errors. This problem can be solved using phase-retrieval algorithms hardly applicable to real time computing [16].

The previous dispersed speckles procedure can easily be adapted to the case of a multi-spectral phase-shifting TI as illustrated in Figure 10. For that purpose the relation (8) should be rewritten as follows:

$$C(P,\sigma) \approx \sum_{n=1}^{N} \exp[2i\pi\sigma\xi_n] \, B_D(P - P_n). \tag{A1}$$

The Fourier transform of Eq. (A1) with respect to the variable σ is defined as:

$$\hat{C}(P,\xi) = \int_{-\infty}^{+\infty} C(P,\sigma)\exp[-2i\pi\sigma\xi]d\sigma. \tag{A2}$$

Combining the two previous relationships readily leads to:

$$\hat{C}(P,\xi) = \sum_{n=1}^{N} B_D(P - P_n) \int_{-\infty}^{+\infty} \exp[-2i\pi\sigma(\xi - \xi_n)]d\sigma, \tag{A3}$$

which, using an elementary property of Fourier transformation, reduces to:

$$\hat{C}(P,\xi) = \sum_{n=1}^{N} B_D(P - P_n)\delta(\xi - \xi_n), \tag{A4}.$$

δ(ξ) being the Dirac distribution. It turns out that the n$^{th}$ sub-pupil will be shifted of a quantity $\xi_n$ along the vertical axis of the transformed data-cube, allowing in principle a fast and accurate determination of its piston error. From a practical point of view however, the Fourier transform along the piston axis must obey to the classical relationship:

$$N_C = 2\, \xi_{Max}/\delta\xi, \tag{A5}$$

where $N_C$ is the total number of spectral channels, and $\xi_{Max}$ and $\delta\xi$ respectively are the half-range and measurement accuracy of the piston errors. Assuming $\xi_{Max}$ = 10 µm and $\delta\xi$ = 50 nm as in the present study we find $N_C$ = 400, an excessive number that would certainly degrade the signal-to-noise ratio of the sensor, and therefore its measurement accuracy. It is worth mentioning, however, that a related method is being currently developed for the PRIMA instrument at the focus of the VLTI, making use of a dramatically reduced number of spectral channels [26].

**FIGURES CAPTIONS**

Figure 1: Basic principle of the phase-shifting TI.

Figure 2: Possible phase shift values $\phi_m$ for cases M = 3 and M = 4.

Figure 3: Flow-chart of the monochromatic OPD retrieval procedure.

Figure 4: Pseudo-code of the OPD unwrapping procedure, with NINT standing for the "Nearest Integer" function.

Figure 5: Possible optical implementation of the piston sensor (not to scale). Plain lines indicate object/image conjugations, while dashed/dotted lines refer to pupil imaging.

Figure 6: Gray-scale map (top) and three-dimensional view (bottom) of the WFE of the segmented mirror affected with random piston and tilt errors (PTV = 15.697 µm and RMS = 4.557 µm).

Figure 7: Top row, pupil transmission map (left) and initial piston errors (right, PTV = 15.463 µm and RMS = 4.555 µm). Middle row: PSF in the image plane (linear and logarithmic scales). Bottom row: reconstructed WFE (left, PTV = 15.484 µm and RMS = 4.555 µm) and comparison with initial errors (right, PTV = 0.071 µm and RMS = 0.014 µm).

Figure 8: OPD estimation error maps for various spectral bandwidths.

Figure 9: Same representations than in Figure 7 with additional seeing perturbations. Top right panel: perturbed WFE for a Fried's radius $r_0$ = 500 mm (PTV = 17.924 µm and RMS = 4.591 µm). Middle left panel: PSF in the image plane (logarithmic scale). Middle right panel: reconstructed WFE (PTV = 18.294 µm and RMS = 4.422 µm). Bottom row: comparison with initial errors (left, PTV = 5.454 µm and RMS = 0.502 µm for the whole 18 segments; right, PTV = 1.403 µm and RMS = 0.151 µm with segment n°10 excluded).

Figure 10: Possible application of the dispersed speckles method to the multi-spectral PSTI.

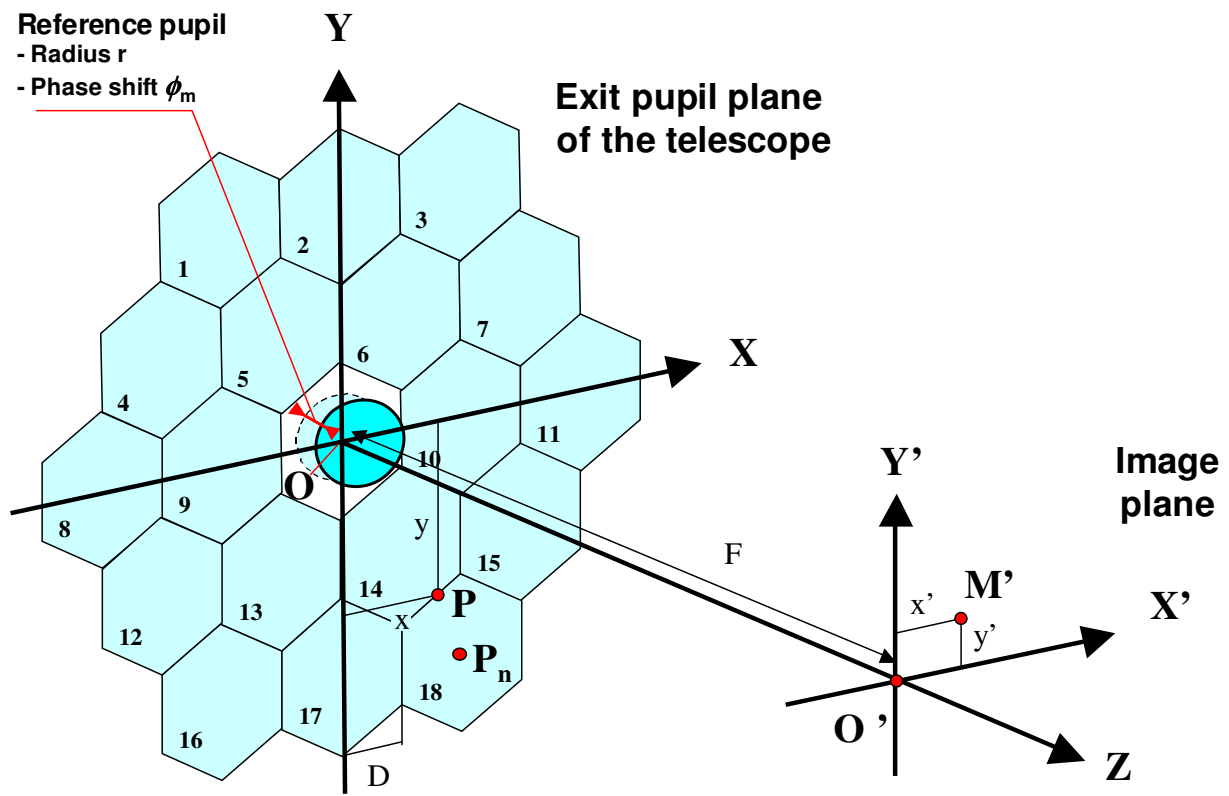

**Figure 1: Basic principle of the phase-shifting TI.**

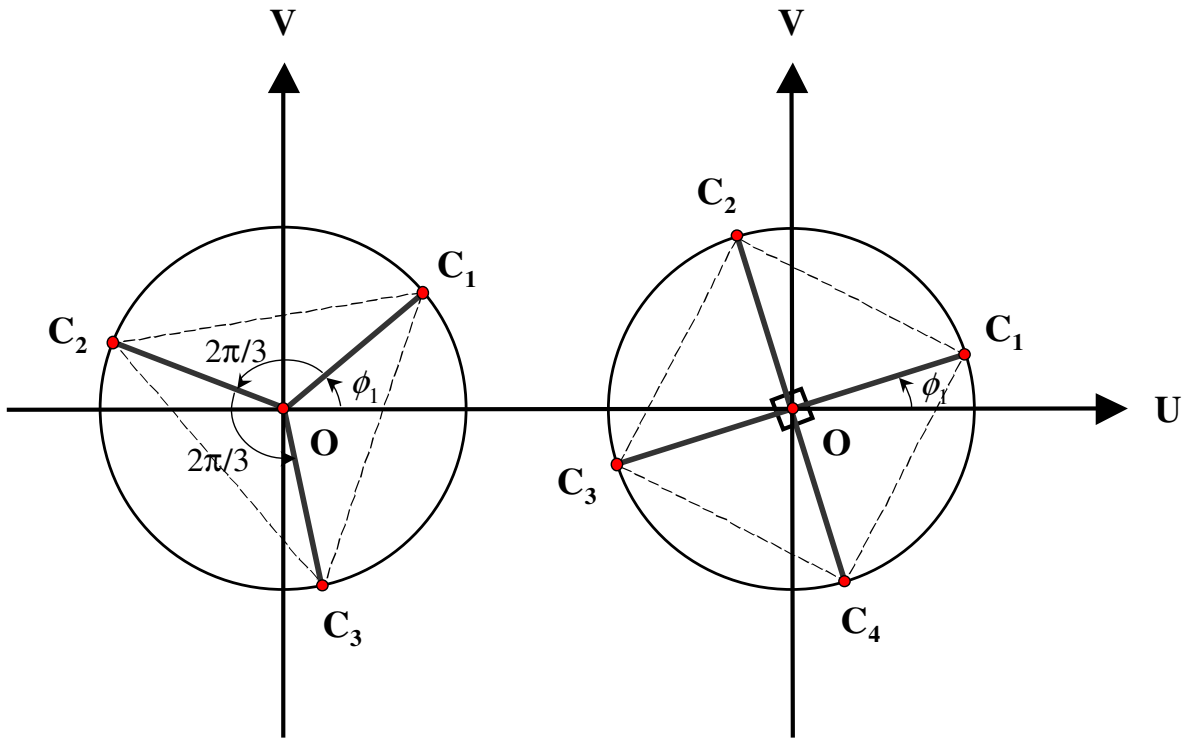

**Figure 2: Possible phase shift values $\phi_m$ for cases M = 3 and M = 4.**

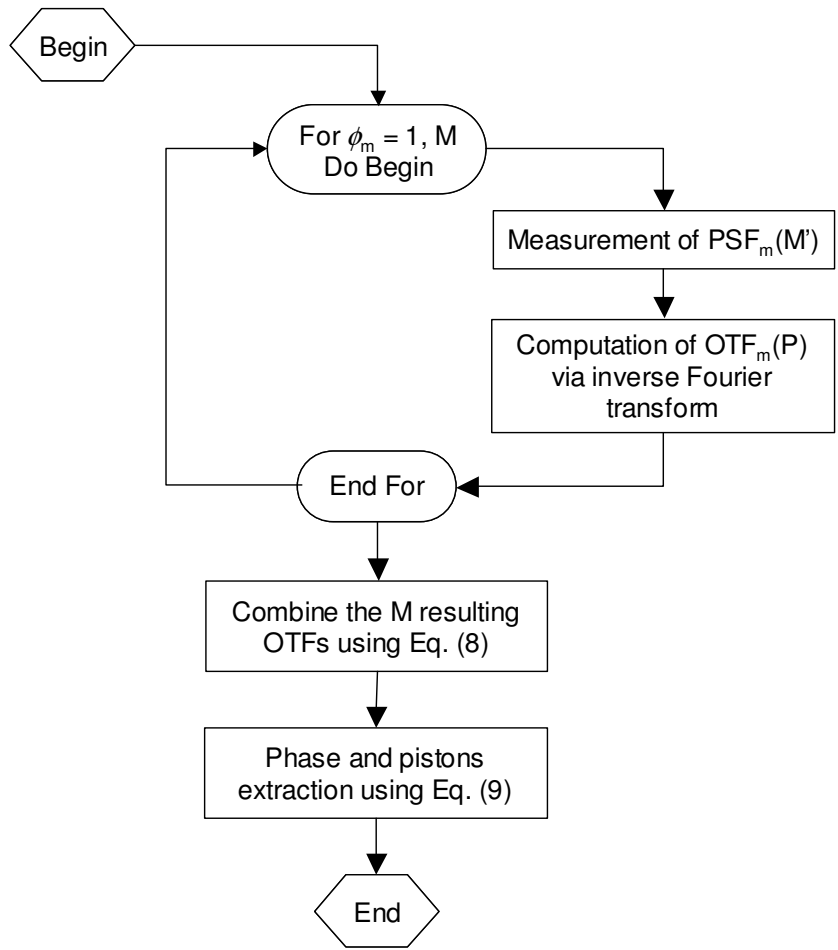

**Figure 3: Flow-chart of the monochromatic OPD retrieval procedure.**

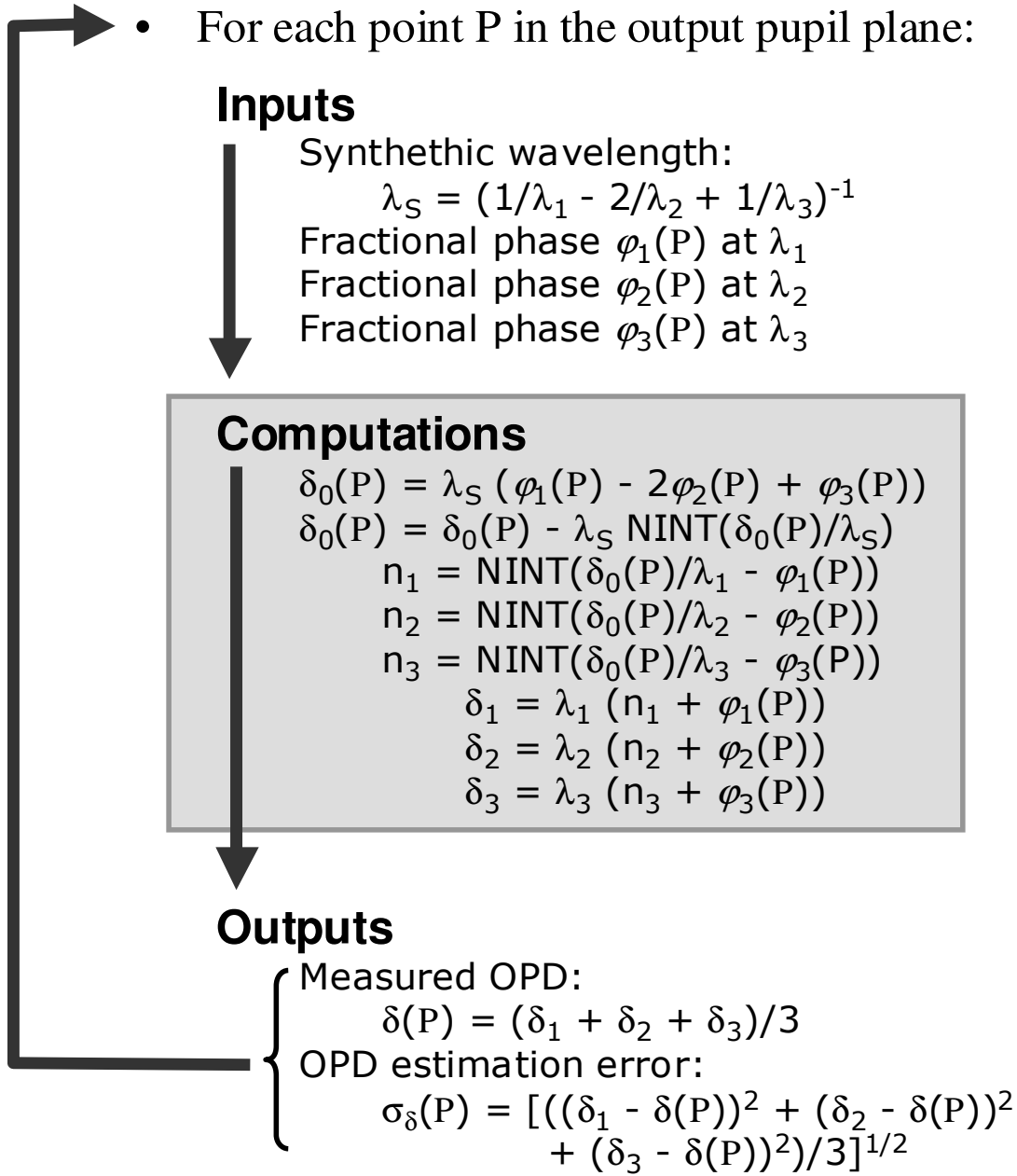

- For each point P in the output pupil plane:

    **Inputs**
    Synthethic wavelength:
    $\lambda_S = (1/\lambda_1 - 2/\lambda_2 + 1/\lambda_3)^{-1}$
    Fractional phase $\varphi_1(P)$ at $\lambda_1$
    Fractional phase $\varphi_2(P)$ at $\lambda_2$
    Fractional phase $\varphi_3(P)$ at $\lambda_3$

    **Computations**
    $\delta_0(P) = \lambda_S (\varphi_1(P) - 2\varphi_2(P) + \varphi_3(P))$
    $\delta_0(P) = \delta_0(P) - \lambda_S \text{ NINT}(\delta_0(P)/\lambda_S)$
    $n_1 = \text{NINT}(\delta_0(P)/\lambda_1 - \varphi_1(P))$
    $n_2 = \text{NINT}(\delta_0(P)/\lambda_2 - \varphi_2(P))$
    $n_3 = \text{NINT}(\delta_0(P)/\lambda_3 - \varphi_3(P))$
    $\delta_1 = \lambda_1 (n_1 + \varphi_1(P))$
    $\delta_2 = \lambda_2 (n_2 + \varphi_2(P))$
    $\delta_3 = \lambda_3 (n_3 + \varphi_3(P))$

    **Outputs**
    Measured OPD:
    $\delta(P) = (\delta_1 + \delta_2 + \delta_3)/3$
    OPD estimation error:
    $\sigma_\delta(P) = [((\delta_1 - \delta(P))^2 + (\delta_2 - \delta(P))^2 + (\delta_3 - \delta(P))^2)/3]^{1/2}$

**Figure 4: Pseudo-code of the OPD unwrapping procedure, with NINT standing for the "Nearest Integer" function.**

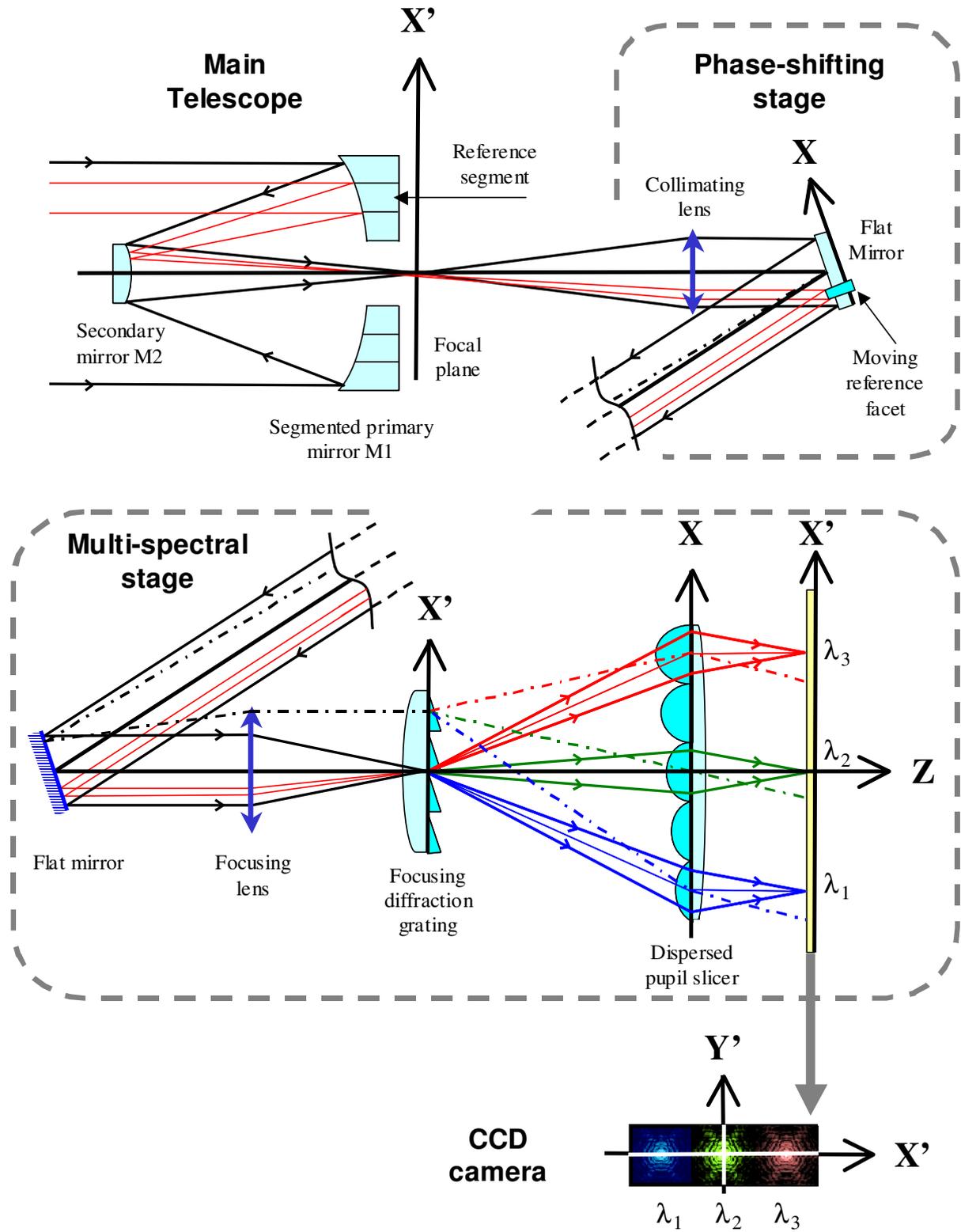

**Figure 5: Possible optical implementation of the piston sensor (not to scale). Plain lines indicate object/image conjugations, while dashed/dotted lines refer to pupil imaging.**

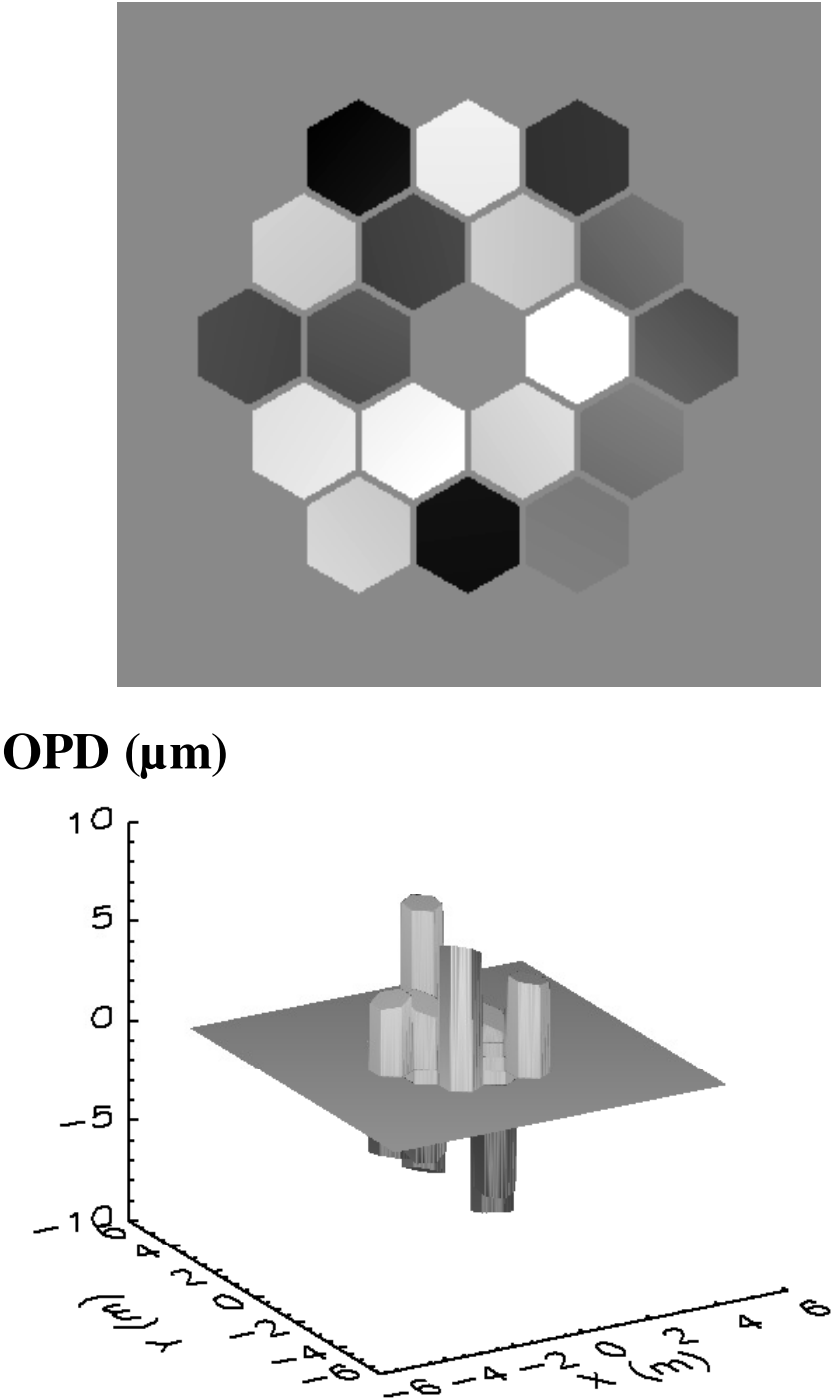

**Figure 6: Gray-scale map (top) and three-dimensional view (bottom) of the WFE of the segmented mirror affected with random piston and tilt errors (PTV = 15.697 μm and RMS = 4.557 μm).**

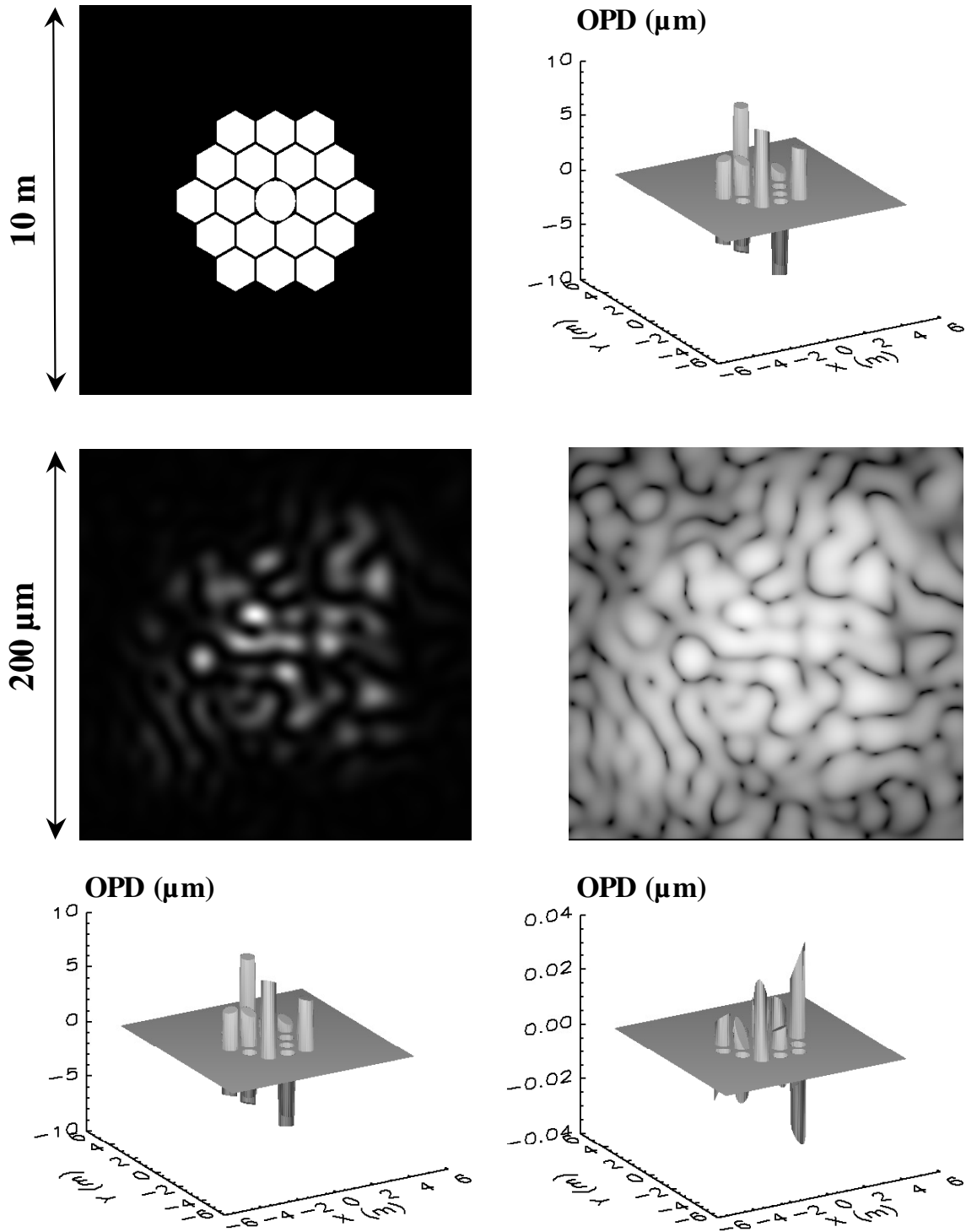

**Figure 7: Top row, pupil transmission map (left) and initial piston errors (right, PTV = 15.463 µm and RMS = 4.555 µm). Middle row: PSF in the image plane (linear and logarithmic scales). Bottom row: reconstructed WFE (left, PTV = 15.484 µm and RMS = 4.555 µm) and comparison with initial errors (right, PTV = 0.071 µm and RMS = 0.014 µm).**

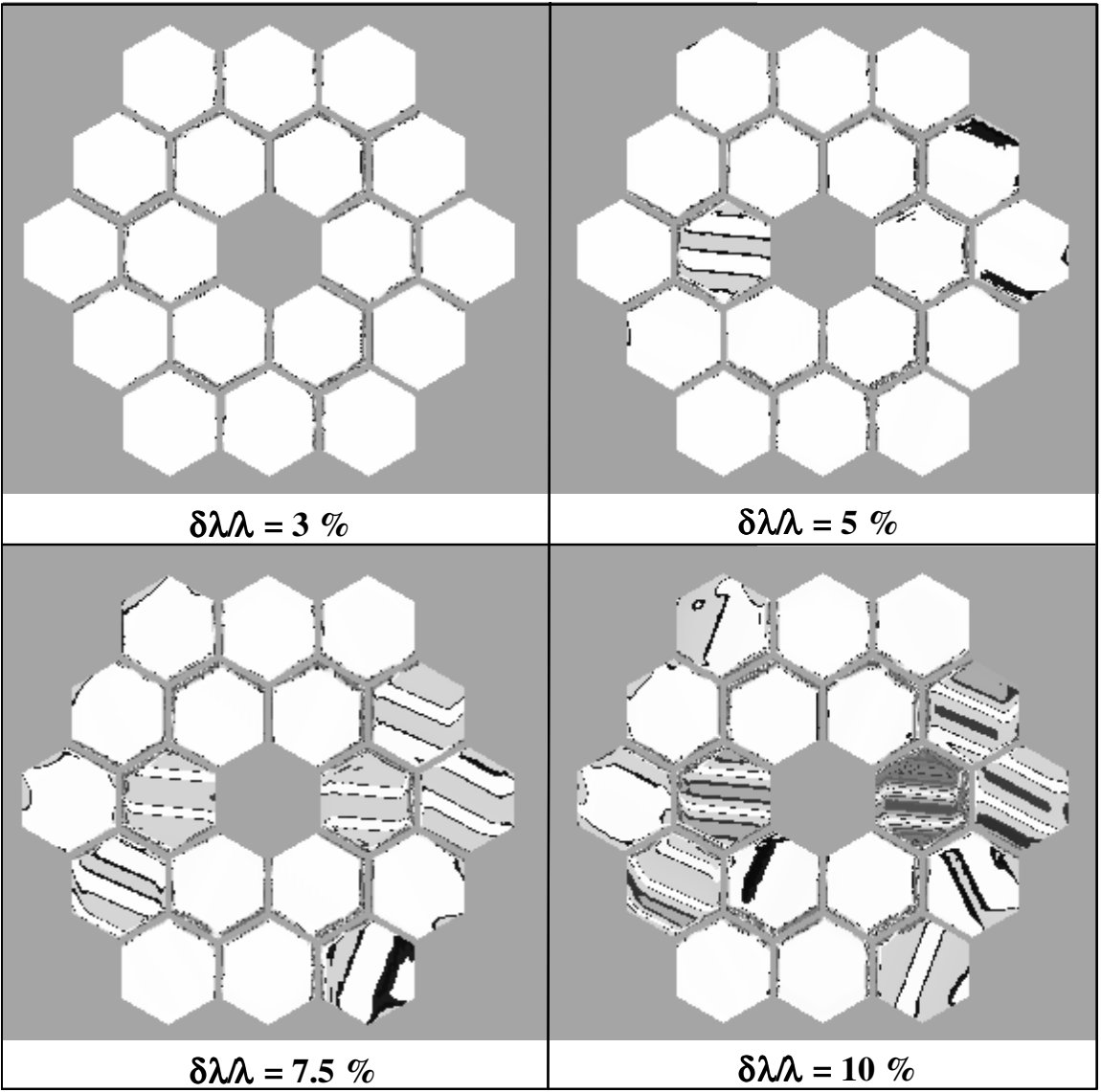

**Figure 8: OPD estimation error maps for various spectral bandwidths.**

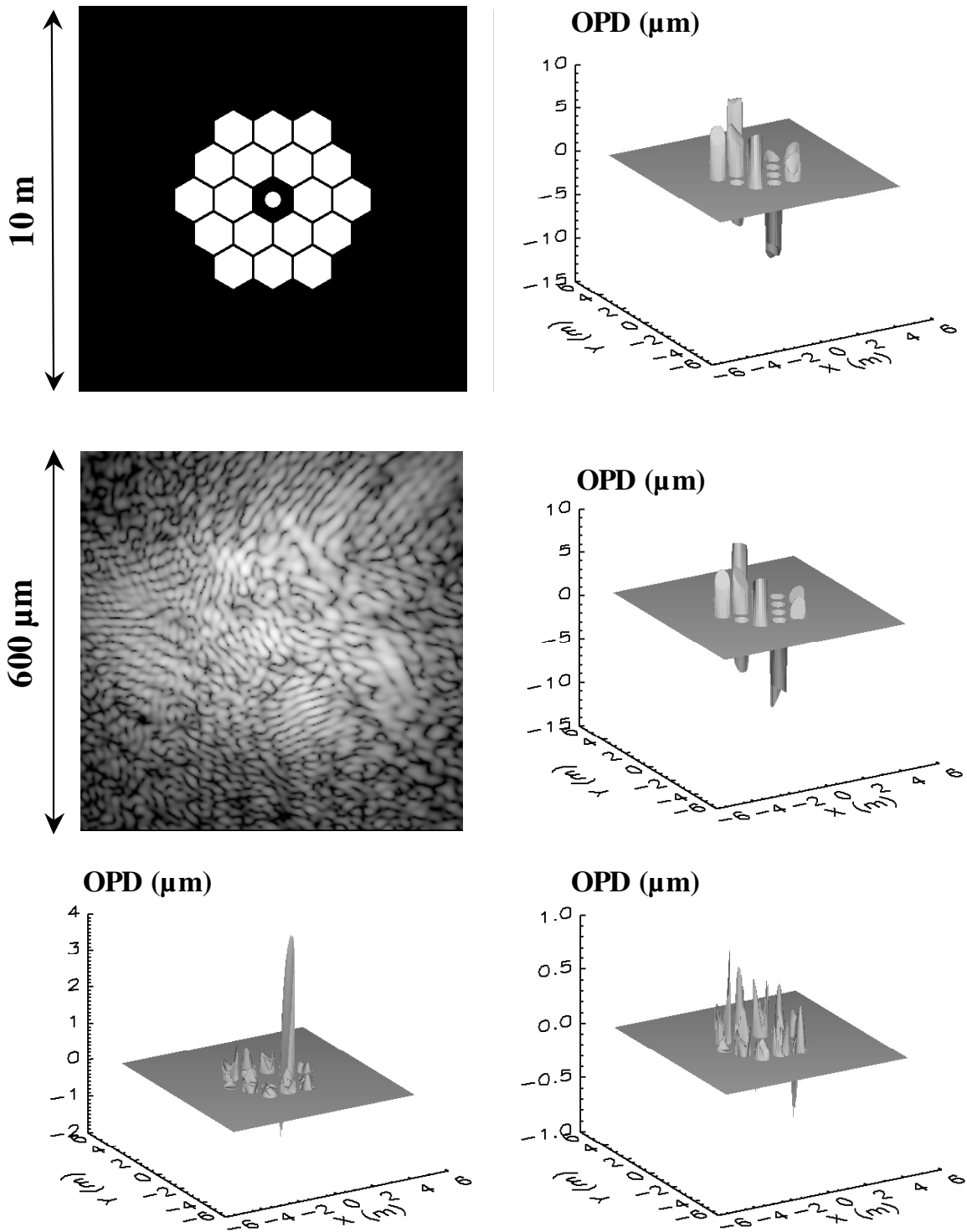

**Figure 9: Same representations than in Figure 7 with additional seeing perturbations. Top right panel: perturbed WFE for a Fried's radius $r_0$ = 500 mm (PTV = 17.924 µm and RMS = 4.591 µm). Middle left panel: PSF in the image plane (logarithmic scale). Middle right panel: reconstructed WFE (PTV = 18.294 µm and RMS = 4.422 µm). Bottom row: comparison with initial errors (left, PTV = 5.454 µm and RMS = 0.502 µm for the whole 18 segments; right, PTV = 1.403 µm and RMS = 0.151 µm with segment n°10 excluded).**

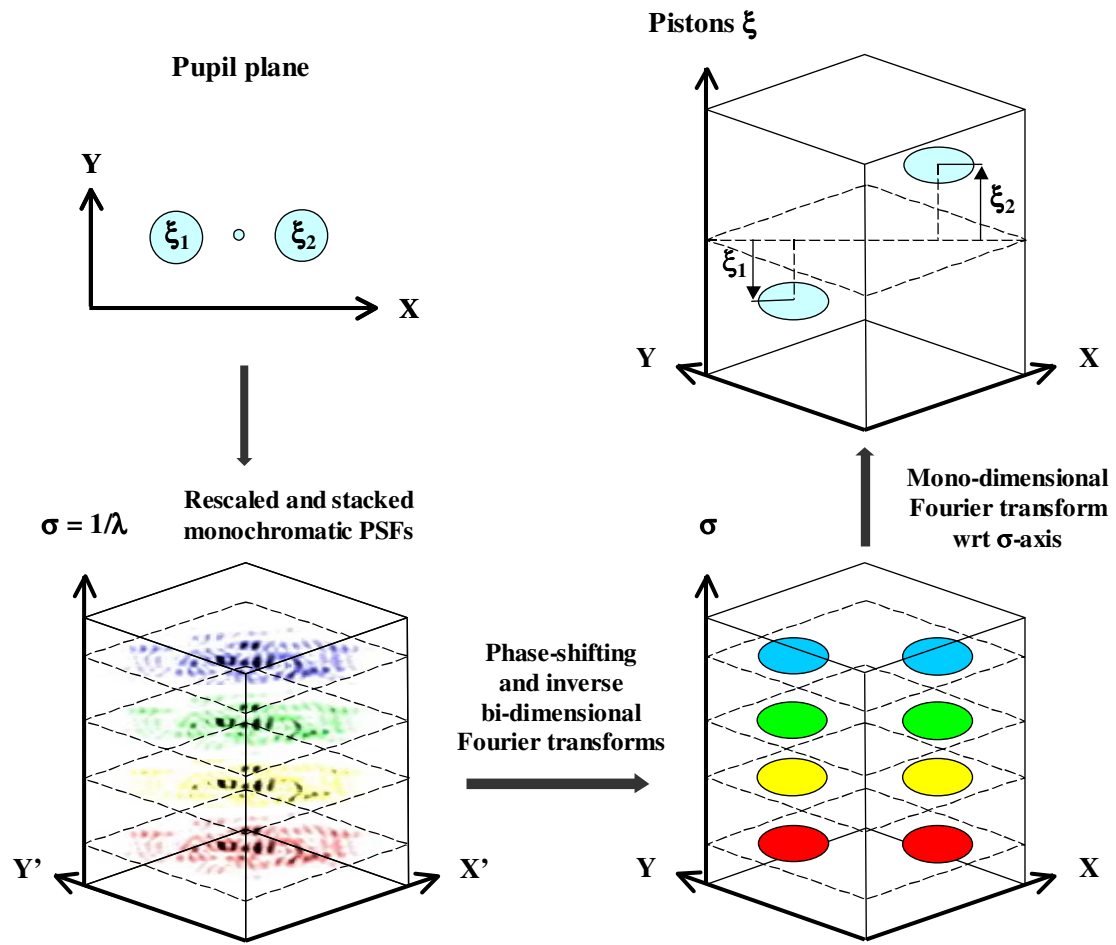

**Figure 10:** Possible application of the dispersed speckles method to the multi-spectral PSTI.